\documentclass[review]{elsarticle}

\journal{Journal of \LaTeX\ Templates}

\usepackage{ stmaryrd }
\usepackage{adjustbox}
\usepackage{graphicx}
\usepackage{amsthm}
\usepackage{amsmath}
\usepackage{graphicx}
\graphicspath{ {images/} }
\theoremstyle{definition}
\newtheorem{definition}{Definition}

\begin{document}

\begin{frontmatter}

\title{An efficient and secure two-party key agreement protocol based on chaotic maps\tnoteref{mytitlenote}}

\author[mymainaddress]{Nahid Yahyapoor}

\author[mysecondaryaddress]{Hamed Yaghoobian}
\cortext[mycorrespondingauthor]{Corresponding author}\ead{hy@uga.edu}

\author[mysecondaryaddress]{Manijeh Keshtgari}

\address[mymainaddress]{Electrical Engineering, Khavaran Institute of Higher Education, Mashhad, Iran}
\address[mysecondaryaddress]{Computer Science, University of Georgia, Athens, GA 30602, USA}

\begin{abstract}
Secure communication is a matter of genuine concern that includes means whereby entities can share information without a third party's interception. Key agreement protocols are one of the common approaches in which two or more parties can agree upon a key, which precludes undesired third parties from forcing a key choice on them. Over the past decade, chaos-based key agreement protocols have been studied and employed widely. Recently, Yoon and Jeon proposed a novel key agreement protocol based on chaotic maps and claimed security and practicality for their protocol. We find that Yoon-Jeon's protocol suffers certain issues: (1) It introduces a trusted third party whose very presence increases the implementation cost. (2) requires a multiplicity of encryption/decryption computations and (3) does not protect the user's anonymity. In order to overcome these problems, we present an enhanced key agreement protocol with user anonymity. Theoretical analysis demonstrates that the proposed protocol is efficient and resists current attacks.  
\end{abstract}

\begin{keyword}
\texttt Session key, Key agreement protocol, Chebyshev chaotic map, Chinese remainder theorem, Security, Anonymity
\end{keyword}

\end{frontmatter}

%\linenumbers

\section{Introduction}
\medskip
Over the last few decades, chaos-based cryptography has been studied extensively. A chaotic system is associated with particular properties such as sensitivity to parameters and initial conditions, pseudo-randomness, and ergodicity. These properties fulfill several certain features such as diffusion and confusion that are required in modern cryptography. The sensitivity to initial conditions and pseudo-randomness of Chebyshev map makes it prominently used in encryption schemes, hash functions, and particularly in key agreement protocols. A key agreement protocol is a protocol in which two or more communication parties create a shared key by using the messages they have sent to one another. Then, this shared key, called a session key, will be used for information encryption/decryption in subsequent communications. Whitefield Diffie and Martin Hellman \cite{diffie1976new} developed and then registered the first key agreement protocol. However, their protocol failed to provide mutual authentication between communication parties, and, therefore, was vulnerable to man-in-the-middle attack. Since then, several key agreement protocols have been designed to prevent man-in-the-middle and related attacks. 

Kocarev and Tasev \cite{kocarev2003public} proposed a public-key encryption scheme based on chaotic maps. Bergamo et al. \cite{bergamo2005security} pointed out that Kocarev-Tasev's presented protocol is insecure, due to the cosine function periodicity, an adversary is able to recover the plaintext from a given ciphertext without any required secret key. Xiao et al. \cite{xiao2007novel} designed a novel key agreement protocol upon which Han in 2008 \cite{han2008security} presented two attacks that  enables an adversary to prevent the user and the server from establishing a shared key. Furthermore, Xiang et al. \cite{xiang2009security} pointed out that Xiao et al.'s protocol is vulnerable to the stolen-verifier attack and the off-line password guessing attack. Later, Han and Chang \cite{han2009chaotic} presented an enhanced protocol, which worked with or without clock synchronization. In 2010, Wang and Zhao \cite{wang2010improved} proposed a modified chaos-based protocol. Yoon and Jeon \cite{yoon2011efficient} proved that Wang-Zhao's protocol requires timestamp information and is vulnerable to illegal message modification attacks. In addition, it has redundant encryption/decryption computations so as to establish a secure key agreement protocol.

It is noteworthy, that none of these protocols is able to protect the anonymity of users over communication channels, whereas in many fields such as electronic commerce, electronic banking and remote Telecare Medicine Information Systems, users should retain their privacy while communicating with the servers. Therefore, in 2009, Tseng et al. \cite{tseng2009chaotic} presented the first key agreement protocol with user anonymity. Later, Niu and Wang \cite{niu2011anonymous} pointed out that it fails to provide user anonymity, perfect forward secrecy, and security against an insider attacker, then proposed a new key agreement protocol. Soon, Yoon \cite{yoon2012efficiency} proved that Niu-Wang's protocol is vulnerable to Denial of Service (DoS) attack and is fraught with computational problems. Tseng and Jou \cite{tseng2011efficient} suggested a key agreement protocol based on chaotic maps, which allows users to interact with the server anonymously. Over the recent years, key agreement schemes using smart cards have received a lot of attention. Das \cite{das2011analysis} proposed a protocol using smart cards, and claimed immunity to attacks. However, Lee and Hsu \cite{lee2012extended} showed it vulnerability to privileged insider attack and off-line password guessing attack and inability to protect the identity of users which resulted in a new modified protocol. Moreover, Lee et al. \cite{lee2012extended} proposed a protocol using smart cards but unfortunately, He et al. proved that Lee et al’s protocol is vulnerable to privileged insider attack, Denial of Service attack, and fails to protect the anonymity of users and as a result proposed a new protocol \cite{he2012cryptanalysis}. 

In this paper, first, we offer a review of Yoon-Jeon's protocol and examine its failure to protect the identity of users while determining a shared session key, redundant encryption/decryption computations and  trusted third party whose presence causes delay, sensitivity, and cost increase in a network. We propose an enhanced key agreement protocol to overcome these problems.

This paper is structured as follows: Section 2 gives a description of the Chebyshev chaotic map, Logistic chaotic map, and the Chinese remainder theorem. In section 3, we study Yoon-Jeon's key agreement protocol. In section 4, we introduce a novel, secure key agreement protocol with user anonymity and then analyze the security and efficiency of the proposed protocol in section 5. Finally, we conclude in section 6.

\section{Preliminaries}
In this section, we introduce some concepts used in our protocol, such as the Chebyshev chaotic map, the Logistic chaotic map, and the Chinese remainder theorem.

\subsection{Chebyshev chaotic map}

\theoremstyle{definition}
\begin{definition}{}
	Let $n$ be an integer and $x$ a variable over the interval $\big[-1,1\big]$. The degree-n Chebyshev polynomial for $x$, is defined using the following recurrence relation:
	\begin{equation}
	T_n(x)=2xT_{n-1}(x)-T_{n-2}(x),
	\label{eq1}
	\end{equation}
where $n\geq2$, $T_0(x)=1$ and $T_1(x)=x$. Some examples of Chebyshev polynomials are:

\begin{equation}
T_2(x)=2x^2-1,
\label{eq2}
\end{equation}	
\begin{equation}
T_3(x)=4x^3-3x,
\label{eq3}
\end{equation}
\begin{equation}
T_4(x)=8x^4-8x^2+1.
\label{eq4}
\end{equation}
	\label{def1}
\end{definition}
\theoremstyle{definition}
\begin{definition}{}
	Let $n$ be an integer and  $x$  a variable over the interval $\big[-1,1\big]$. The polynomial $T_n(x)=\big[-1,1\big]\shortrightarrow \big[-1,1\big]$, is used as:
	\begin{equation}
	T_n(x)=\cos(n\arccos(x)),
	\label{eq5}
	\end{equation}	
	\label{def2}
\end{definition} 
Definitions \ref{def1} and \ref{def2} are equivalent. Chebyshev Polynomials have two important properties, they are semi-group and chaotic.
\theoremstyle{definition}
\begin{definition}{}
	The semi-group property: One of the most important properties of Chebyshev polynomials is the semi-group property, which is defined with:
	\begin{equation}
	T_r(T_s(x))=T_s(T_r(x))=T_{rs}(x),
	\label{eq6}
	\end{equation}	
	\label{def3}
\end{definition}
 \begin{definition}{}
 	The chaotic property: If the degree $n>1$, Chebyshev polynomial map $T_n(x)=[-1,1] \shortrightarrow [-1,1]$ is a chaotic map with invariant density $f(x)=\frac{1}{\pi\sqrt{1-x^2}}$ for positive Lyapunov exponent $\lambda=\ln n$.
 	\label{def4}
 \end{definition}
\begin{definition}{}
	Enhanced Chebyshev polynomial: Zhang \cite{zhang2008cryptanalysis} proved that the semi-group property holds true for Chebyshev polynomials in the interval $(-\infty,+\infty)$. Enhanced Chebyshev polynomials are defined as:
	\begin{equation}
	T_n(x)=2xT_{n-1}(x)-T_{n-2}(x)(\mathrm{mod}(N)),
	\label{eq7}
	\end{equation}	
	where, $n\geq2$, $x\in(-\infty,+\infty)$ and $N$ is a large prime number.
	\label{def5}
\end{definition}
 \begin{definition}{}
	The Diffie-Hellman problem (DHP): DHP is defined as: two different degree polynomials $T_r(x)$ and $T_s(x)$ are assumed, finding $T_{rs}(x)$ is impossible without knowing $r$ and $s$.
	\label{def6}
\end{definition}
 \begin{definition}{}
	The discrete logarithm problem (DLP): DLP is defined as: an element $a$ is assumed, finding the integer $r$ so that $T_r(x)\equiv a$ is impossible. 
	\label{def7}
\end{definition}
\subsection{Logistic chaotic map}
One of the simplest chaotic maps is the Simple Logistic Function (SLF). It can be expressed as:
\begin{equation}
x_{n+1}=\lambda x_n(1-x_n),
\end{equation}
where, $n=0, 1, 2, 3,\cdots$ , $x_0\in[0,1]$  is an initial value, $x_n$ the $n$th value in the sequence, accordingly, $x_{n+1}$ the ${n+1}$th term in the same sequence and $0\leq\lambda\leq4$ the logistic map parameter. When we adjust the $\lambda$ parameter beyond 3.57, we see the onset of chaos. In fact, for a behavior to be chaotic, $\lambda$ should be between 3.57 and 4.
\subsection{Chinese remainder theorem}
The Chinese remainder theorem or CRT for short, has been employed vastly in cryptography. This algorithm hides data and is hypothetically designed as a one-way function. The theorem is described as: suppose $m_1,m_2,\cdots,m_r$ are positive integers that are pairwise co-prime numbers and $a_1,a_2,\cdots,a_r$ is the sequence of the given integers where: 
\begin{equation} \label{alarm}
\left\{
\begin{array}{ll}
x\equiv a_1 \mod m_1 \\
x\equiv a_2 \mod m_2 \\
\vdots\\
x\equiv a_r \mod m_r\\
\end{array}
\right.
\end{equation}

Then $x \overset{m_i}\equiv a_i$, $i=1,2,3,\cdots,r$ has only one answer to the module $M=\prod_{i=1}^{r} m_i$ and equals:
\begin{equation}
X=\sum_{i=1}^{r}a_iM_iy_i
\end{equation}
where, $M_i=\frac{M}{m_i}$, $y_i=M_i^{-1}\mod m_i$
\section{Yoon-Jeon's key agreement protocol based on Chebyshev chaotic map}
This section reviews the Yoon-Jeon protocol \cite{yoon2011efficient}. All the notations used in the Yoon-Jeon protocol are described in Table 1. Assume Alice and Bob are two participants in a key agreement process. In this system, Trent is a trusted third party in the network, e.g., KDC (key distribution center) which publishes the system parameters including Chebyshev polynomials, E(.), D(.), and H(.) prior to the commencement of key agreement protocol and also shares a different secret key with each participant. The protocol is as follows:
\begin{enumerate}
\item Alice selects a large integer $r$, a large prime number $N$, and a random number $x\in(-\infty,+\infty)$, and then computes $T_r(x)$, where, $T_r(x)$ is a n-degree Chebyshev polynomial in $x$. She concatenates $A$, $B$, $x$, $N$ and $T_r(x)$.
\item Trent decrypts $E_{TA}(A,B,x,N, T_r(x))$ and checks whether $A$ is a valid identity. If not, Trent stops here; otherwise, Trent concatenates $B$, $A$, $x$, $N$ and $T_r(x)$ and encrypts them using the shared key with Bob and sends $E_{TB}(B,A,x,N,T_r(x))$ to Bob.
\item Having received the message, Bob decrypts the cipher-text and checks whether $B$ is his identity. If not, Bob stops here; otherwise, he selects a large integer $s$ and computes $T_s(x)$), the shared session key $k=T_s(T_r(x))$, and the authentication value $MAC_B=H_k(B,A,T_r(x))$. Bob sends $T_s(x),MAC_B$ to Alice.
\item Alice computes the shared session key $k=T_r(T_s(x))$ and the same authentication value $MAC_B^\prime=H_k(B,A,T_r(x))$. Then, she checks whether $MAC_B$ and $MAC_B^\prime$ are equal. If so, the Bob's identity is authenticated. Next, Alice calculates the authentication value $MAC_A=H_k(B,A,T_s(x))$ and sends it to Bob.
\begin{table}
\caption{Notations used in Yoon-Jeon's protocol}
\begin{center}
	\begin{tabular}{|c c|} 
		\hline
		Symbol & Definition \\ [0.5ex] 
		\hline\hline
		$A,B$ & Identifiers of Alice and Bob, respectively \\ 
		\hline
		$TA$, $TB$ & Shared secret key between Alice and Bob with Trent, respectively \\
		\hline
		$T_n(x)$ & Chebyshev polynomial in $x$ of degree $n$ \\
		\hline
		$E(.)$ & A symmetric encryption algorithm \\
		\hline
		$D(.)$ & An asymmetric decryption algorithm \\ 
		\hline
		$H(.)$ & A one-way hash function\\
		\hline
		$K$ & Finally established session key between Alice and Bob\\ [0.5ex] 
		\hline
	\end{tabular}
\end{center}
\end{table}
\item Having received the message, Bob computes $MAC_A^\prime=H_k(A,B,T_s(x))$ and checks whether $MAC_A$ and $MAC_A^\prime$ are equal; if so, the identity of Alice is authenticated.

Therefore, Alice and Bob have achieved the shared session key $k=T_r(T_s(x))=T_s(T_r(x))=T_{rs}(x)$ in order to protect the exchanged information in subsequent communications.

An absolute trust in the key distribution center is presumed in the above protocol. Private keys are issued by the key distribution center for the server and users. Nevertheless, the possibility of encrypted message abuse by the center exists, and it is clear that determining a trusted third party is difficult. Each complete execution of the key agreement protocol requires two encryptions and two decryptions. Also in the first step of the protocol, the adversary can obtain the user's ID, thereby compromising the true identity of the user. 
\end{enumerate}

\section{The proposed protocol}
In this section, we introduce an efficient and secure key agreement protocol based on chaotic maps which protects the users' anonymity. It incorporates two phases of registration and authentication-key agreement. The notations used in our protocol are listed in Table 2.
\begin{enumerate}
	\item Registration phase
	\begin{enumerate}
		\item $U_i$ chooses a large integer $n$, a random parameter $\lambda$ over the interval $\big[3.57, 4\big]$ and an initial value $x_0$ over the interval $\big[0,1\big]$ so as to generate the chaotic sequence $A=(x_0,x_1,\dots,x_n)$ using Logistic mapping. She also selects a positive integer $m_1$, a password $pw_i$ and a random nonce $n_s$. Now, she sends $\big\{ID_i,a,m_1,h_{pw}\big\}$ to the server over a secure channel, where $ID_i$ is the identity, $a$ sum of all the elements in $A$ and $h_{pw}=H(pw_i, n_s)$. 
		
		\begin{table} [!h]
			\caption{Notations used in our proposed protocol}
			\begin{center}
			\begin{adjustbox}{max width=\textwidth}
				\begin{tabular}{|c | c|} 
					\hline
					Symbol & Definition \\ [0.5ex] 
					\hline\hline
					$U_i$ & Some user $i$ \\ 
					\hline
					$ID_i$, $ID_s$ &  Identities of the user $i$ and the server \\
					\hline
					$pw_i$ & A password chosen by user $i$ \\
					\hline
					$n_s$, $k$ & Nonces chosen by user $i$ \\
					\hline
					$M_i$ & Nonces chosen by the server \\ 
					\hline
					$T_n(x)$ & The Chebyshev polynomial in $x$ of degree $n$\\
					\hline
					$H(.)$ & A one-way hash function\\
					\hline
					$sk_i$ & Finally established session key between the user $i$ and the server\\
					\hline 
					$\oplus$ & The exclusive-or operation\\ [0.5ex] 
					\hline
				\end{tabular}
            	\end{adjustbox}
			\end{center}
		\end{table}
		
		\item Similar to the server $U_i$ chooses a large integer $n^\prime$, a random parameter $\lambda^\prime$ over the interval , an initial value $x_0 ^\prime$ over the interval $\big[0, 1\big]$ so as to generate the chaotic sequence $B=\big(x_0 ^\prime, x_1 ^\prime, \dots, x_n ^\prime\big)$ using logistic mapping. Having acquired $b$ which is the sum of elements in the chaotic sequence $B$, the server proceeds to obtain $b^\prime$ and $a^\prime$ by multiplying $a$ and $b$ by $10^c$ and $10^{c^\prime}$ where $c$ and $c^\prime$ are the decimal digits of $a$ and $b$. Furthermore, the sever chooses the positive integer $m_2$ as mutually prime to $m_1$. The server now uses the Chinese remainder theorem to calculate $X$ and having acquired $R_i=H\big(ID_i, H_{pw}\big)\oplus H\big(X\big)$ and $R_1=H\big(m_2,h_{pw}\big)$, chooses the random sequence $M_i$ and transmits the message $\big\{ID_s,M_i,R_i,R_1\big\}$ over a secure channel to the user with $ID_s$ identity. The registration phase is illustrated in Figure 1 below. 
	\end{enumerate}
\begin{figure}[!h]
	\begin{center}
		\begin{adjustbox}{max width=\textwidth}
	\includegraphics {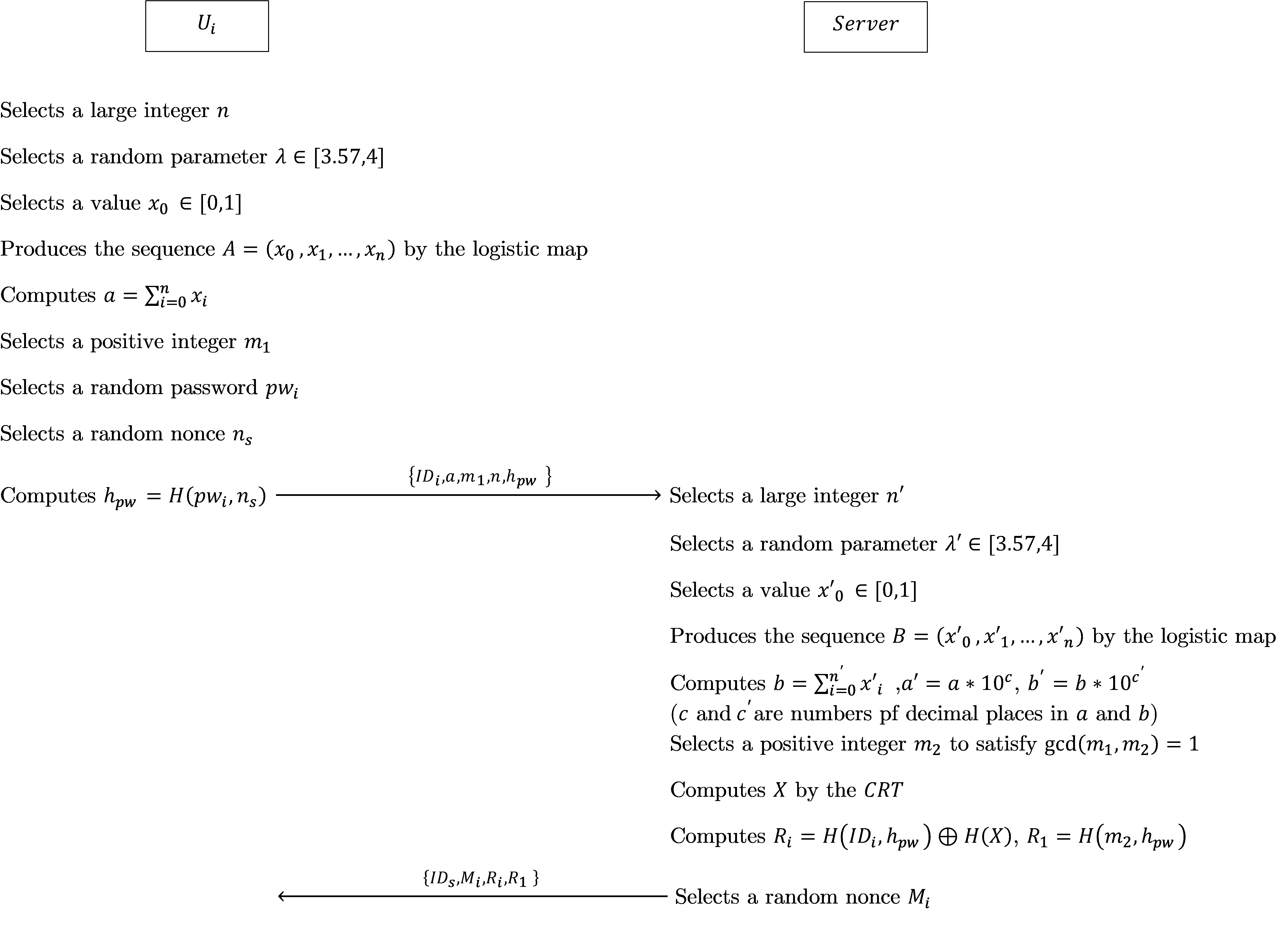}
	\centering
	   	\end{adjustbox}
\end{center}
		\caption{Registration phase of our protocol }
\end{figure}

\item Authentication-key agreement phase 
\begin{enumerate}
	\item $U_i$ selects a large integer $r$, a large prime $N$, and a random number $x\in \big (-\infty,+\infty \big)$, and then computes $T_r(x)$, a $n$-degree Chebyshev polynomial in $x$. Then she computes the parameters below by choosing a random nonce $k$.
	 \begin{equation*}
	 M_1=R_i \oplus H(k)
	 \end{equation*}
	 \begin{equation*}
	 AID_i=ID_i \oplus H(k)
	 \end{equation*}
	 \begin{equation*}
	 M_2=T_r(x) \oplus R_1
	 \end{equation*}
	 Finally, she transmits $\big\{M_i,M_1,M_2,AID_i,x,N\big\}$ to the server. 
	 \item Upon receiving this message and seeing $M_i$, the server searches for the registered user to which this random nonce has been assigned. The server finds $R_i$ and $R_1$ by receiving $M_1$ and $AID_i$, computes $H(k)=R_i \oplus M_1$ and $ID_i^\prime=H(k)\oplus AID_i$. Furthermore, it examines if $ID_i \stackrel{?}{=} ID_i^\ast$ is true or not and in case it is, it establishes the authenticity of the user, proceeding to acquire $T_r(x)=M_2\oplus R_1$ and this time, chooses a large integer $s$ so as to calculate $T_s(x)$, the session key $sk_i=T_s\big(T_r(x)\big)$ and authentication value $AU_s=H\big(ID_i,H(k),sk_i\big)$. Finally, the server transmits the message $\big\{ID_s,M_3,AU_s\big\}$ to the user which is $M_3=T_s(x)\oplus H(x)$.
	 \item The user obtains $H(x)=R_i \oplus H\big(ID_i,h_{pw}\big)$ and $T_s(x)=M_3 \oplus H(x)$ which enables him/her to compute the session key $sk_i=T_r(T_s(x))$ and the server authentication value ${AU_s}^\prime=H\big(ID_i, H(k),sk_i\big)$ by using the previously acquired $T_s(x)$ and possessing $T_r(x)$, which was calculated in the first step of the authentication-key agreement phase. She examines the accuracy of $AU_s^\prime\overset{?}{=}AU_s$ and the correct case establishes the identity of server. Then, she computes her own authentication value $AU_i=H\big(ID_s, H(k),sk_i\big)$.
	 
	 \item Having received the message, the server computes $AU_i\prime=H\big(ID_s, H(k),sk_i\big)$ and examines the accuracy of ${AU_i}^\prime\overset{?}{=}AU_i$, and if it is accurate, the identity of user $i$ is established. Now that the mutual authentication between the user and the server is established, the key $sk_i=T_r(T_s(x))=T_s(T_r(x))$ is used as the shared secret key between these two participants.
\end{enumerate}
\end{enumerate}

\section{Analysis of the proposed protocol}
The performance and security of the proposed protocol is now studied. The theory analysis demonstrates that the offered key agreement protocol is secure and efficient.
\begin{enumerate}
	\item Security analysis 
	\begin{enumerate}
		\item Bergamo et al.'s attack \cite{bergamo2005security}

This attack is possible under two conditions. First, an attacker is to acquire the related parameters $x$, $T_r(x)$ and $T_s(x)$, second, if several Chebyshev polynomials cross the same crossing point, due to the periodicity of cosine functions, the adversary would be able to recover the encrypted text. In the proposed protocol, $T_r(x)$ and $T_s(x)$ are substituted within $M_2=T_r(x) \oplus R_1$ and $M_3=T_s(x) \oplus H(x)$, respectively. Adversaries are not able to acquire these polynomials without knowing $R_1$ and $H(x)$ which are transferred to the user over a secure channel. Besides, the enhanced Chebyshev polynomials employed in this protocol render Bergamo et al.'s attack impossible. 
		\item Man-in-the-middle attack

In the suggested protocol, an adversary cannot forge authentic messages, because users and the server analyze received messages during protocol performance. Then, in the third step of the authentication-key agreement phase, user verifies the authenticity of the server $AU_s=H(ID_i,H(k),sk_i)$, and next, the server verifies $AU_i=H(ID_s,H(k),sk_i)$. Therefore, our protocol is able to prevent these forms of attack. The authentication-key agreement phase is illustrated in figure 2 below. 
		\item Replay attack

		It is a form of network attack, in which a valid data transmission is maliciously or fraudulently repeated or delayed. This is carried out either by the originator or by an adversary who intercepts the data and retransmits it. The adversary eavesdrops on the conversation and having acquired the necessary information such as the user-name and password from the session, sends the password (or hash). For instance, supposing that the user $U_i$ wants to prove his/her identity to the server, the server requests the password as a proof of identity, which the user dutifully provides probably after some transformation like a hash function; meanwhile an adversarial third party is eavesdropping on their conversation and keeps the password. After the interchange is over, the adversary (posing as the user) connects to the server; and when asked for a proof of identity, sends the user's password (or hash) read from the last session, which the server accepts thus granting access to the adversarial third party. Ways to avoid replay attacks include using one-time passwords, (pseudo-) randomly generated strings (nonce), and time-stamping.

	In our key agreement protocol's authentication step, to prevent such attacks, the ends of the communication system are supposed to confirm parts of the received messages. Second, the server acquires $ID_i ^\ast=H(k) \oplus AID_i$ and confirms the authenticity of the user. Third, the user $U_i$, after receiving the message $\big\{ID_s,M_3,AU_s\big\}$ identifies the server and having found $R_i$ in its database, obtains $H(X)=R_i \oplus H \big(ID_i, h_{pw}\big)$ , $T_s=M_3 \oplus H(x)$ and $sk_i T_r(T_s(x))$, thus enabled to confirm the authentication key. Then, the server confirms $AU_i=H(ID_s, H(k), sk_i)$ and all the communicated messages are different because the numbers and random strings $R$, $S$ and $K$ are rapidly changing in the protocol, therefore the adversary fails in any attempt.

		This protocol is vulnerable to attack, mainly because no solution has been offered for authenticity assurances that would affirm the message’s origin.  In the first step of the authentication-key agreement phase of the proposed protocol, $M_1$ , $M_2$ and $AID_i$ are different in each execution, because random nonce $k$ and random number $r$ are chosen by the user in each execution. In the second and third steps, $M_3$, $AU_s$ and $AU_i$ are chosen differently in each execution because the random numbers $r$, $s$ and $sk_i$, $H(k)$ are refreshed. As a result, the replay attack does not work.

	\item Mutual authentication

	In the protocol's authentication-key agreement's thrid step, the server's identity is verified by examining the equality of $AU_s ^\prime$ and $AU_s$ , because only one authorized server is able to compute $AU_s$. Furthermore, in forth step, the server verifies the user by examining the equation ${AU_i}^\prime\overset{?}{=}AU_i$. Because only the authorized user is capable of computing ${AU_i}$. Thus, the server and user reach mutual authentication.
	\item Perfect forward secrecy

	This property of key agreement protocols ensures that compromise of long-term keys does not compromise past session keys.
	It protects past sessions against future compromises of secret keys. The session keys are dependent upon random numbers $r$ and $s$, which are inaccessible to adversaries. The protocol is seen as an important security feature.
	\item Known session key secrecy

The key agreement protocols are supposed to be dynamic so that each execution results in a unique session key. This feature ensures that if an adversary could access a session key, she would be unable to recover the other session keys. Thus, supposing an adversary could obtain a secret session key between the user and server, she would not be able to compute the other session keys because the adversary would face $DHP$ and $DLP$. In addition, the created session keys are chosen by the user and the server, dependent upon random numbers of $r$ and $s$, so they would be different in each protocol execution. Thus, the inability of an adversary to gain random numbers of $r$ and $s$, makes the session keys unattainable. Clearly, the proposed protocol satisfies this need. 
\item Privileged insider attack

Adversaries frequently guess weak passwords using password cracker applications, which make attempts at guessing the passwords by using particular algorithms and keyword dictionaries. One of the most common and biggest mistakes is choosing one password for all accounts. In the first step of the proposed protocol, the user $i$ chooses a random nonce $n_s$ while computing $h_{pw}= H(pw_i, n_s)$, transmits it to the server, and $h_{pw}$ is substituted in $R_1=H(m_2,h_{pw})$ from the second step of the registration phase. It is impossible for a malicious server to guess the $h_{pw}$ password without identifying the random nonce $n_s$, even if the user selected a weak password $pw_i$ that is easier to remember. 
\item User anonymity

In insecure environments such as e-commerce, e-banking, and telecare medicine information systems, when the users intend to agree upon a mutual key session with the server, they also wish to remain anonymous. Therefore, protecting the privacy of users is crucial in such environments, and the key agreement protocols are supposed to be designed in a way to make it impossible for an adversary to extract the identity of users by eavesdropping on conversations between the users and the server. Supposing that an adversary was capable of eavesdropping on all the transmitted messages, he would be able to learn about the true identity of the user. During transmission of the first message in the authentication-key agreement phase, even if the adversary eavesdrops on the message $\big\{M_i,M_1,M_2,AID_i,x,N\big\}$, he still will not be able to obtain $ID_i$ because it is integrated into the equation $AID_i=ID_i \oplus H(k)$, and the adversary is faced to deal with the one-way function $H(k)$ for which he would not have the adequate time to break. Furthermore, $k$ is a random string, changing in each execution of the protocol. In the second message $\big\{ID_s, M_3, AU_s\big\}$, $ID_i$ is placed in the equation $AU_s=H\big(ID_i,H(k),sk_i\big)$ in which the adversary has to break the impenetrable one-way function, therefore the true identity of the user remains protected and anonymous. The adversaries are unable to access the true identity of users, because their identities are substituted in $AID_i=ID_i \oplus H(k)$ and $AU_s=H\big(ID_i,H(k), sk_i\big)$. Since the random nonce $k$ is chosen by users in each protocol execution, adversaries are not capable of guessing a big random nonce and are faced with a one-way hash function so $ID_i$ is beyond access, and as a result the anonymity of users is protected.
\item Server impersonation by insider users

In this proposed protocol, the server, having received $a$ and $M_1$ from the user $i$, chooses $b$ and $m_2$ proceeds to compute $H(x)$ by using the Chinese remainder theorem. $H(x)$ is not the long-term key of all users and is considered unique for each user in the registration phase, making it impossible for insiders to access the other users' $H(x)$, preventing them from introducing themselves as the server.

\end{enumerate}
\item Performance analysis

In this subsection, we compare the suggested protocol with the other protocols which have been offered recently. The number of performed operations in this protocol perfectly accounts for the amount of computation and the required amount of time for its execution. The less is the computation time is, the shorter the protocol execution. Therefore, the protocol with lesser computational complexity is more practical. A comparison of the computation time of the proposed protocol with the other associated protocols are provided in table 3. The parameters below are given for easier performance evaluation.

$T_H$: Computation time of hash function

$T_X$: Computation time of XOR

$T_E$: Computation time of symmetric encryption algorithm

$T_D$: Computation time of symmetric decryption algorithm

$T_{CM}$: Computation time of Chebyshev chaotic map

In regard to the information in table 3, N/A refers to not requiring a trusted third party and the computation time of operators are calculated based on the execution time of the hash function. The time complexity of a XOR operation in comparison with that of hash function can easily be disregarded; the other costs calculated $T_E\approx2.5T_H, T_D\approx2.5 T_H, T_{CM}\approx175T_H$ \cite{fan2010provably}. The number of computational executions (encryption and decryption computations, Chebyshev chaotic map, Hash and XOR functions) in key agreement level for the proposed algorithms is calculated. In order to conduct a practical analysis to make a reasonable comparison between protocols, it is required that all operational units are presented as one, therefore, the operational complexity of units is  calculated according to the Hash function.

Based on the table, it is clear that the amount of computations in the proposed protocol in compariosn to related protocols has decreased, and there is no need for encryption and decryption computations, which resulting in efficiency improvement. 

\begin{table}[!h]
	\caption{Performance comparison of key agreement protocols}
	\begin{center}
		\begin{adjustbox}{max width=\textwidth}
		\begin{tabular}{|c | c | c| c | c |} 
			\hline
			Proposed schemes & Cost of each user & Cost of trusted third party & Cost of the server & Total time \\ [0.5ex] 
			\hline\hline
			Tseng-Jon \cite{tseng2011efficient} & $T_X+3T_H+2T_{CM}+T_E+T_D$ & $T_X+T_H+T_{CM}+2T_E+2T_D$ & $2T_H+T_{CM}+T_E+T_D$ & $726T_H$ \\ 
			\hline
			Niu-Wang \cite{niu2011anonymous} &  $2T_H+2T_{CM}+T_E+T_D$ & $2T_E+2T_D$ & $2T_H+T_{CM}+T_E+T_D$ & $724T_H$ \\
			\hline
			Yoon-Jeon \cite{yoon2011efficient} & $2T_H+2T_{CM}+T_E$ & $T_E+T_D$ & $2T_H+T_{CM}+T_D$ & $714T_H$ \\
			\hline
			He et al \cite{he2012cryptanalysis} & $2T_X+4T_H+3T_{CM}$ & $N/A$ & $3T_X+4T_H+3T_{CM}+T_D$ & $958T_H$ \\
			\hline
			Lee et al \cite{lee2012extended} & $6T_X+7T_H+2T_{CM}$ & $N/A$ & $6T_X+5T_H+2T_{CM}$ & $712T_H$ \\ 
			\hline
			Lee-Hsu \cite{lee2013secure} & $5T_X+10T_H+3T_{CM}$ & $N/A$ & $3T_X+7T_H+3T_{CM}$ & $967T_H$\\
			\hline
			Our proposed protocol & $5T_X+4T_H+2T_{CM}$ & $N/A$ & $4T_X+2T_H+2T_{CM}$ & $706T_H$\\
			\hline
		\end{tabular}
	\end{adjustbox}
	\end{center}
\end{table}

It is understandable from the security analysis that our proposed protocol does not suffer from complexity issues in encryption/decryption operations due to using XOR and Hash functions along with the Chebyshev chaotic map. Other protocols utilize encryption in transmission of their messages. Each encryption/decryption computation includes several Hashing and XOR operations. It is clear that all such protocols are afflicted with complexity issues and require more execution time than the proposed protocol. 

\end{enumerate}

\section{Conclusion}

Since reaching to a certain satisfying level of security with minimum computations in designing a protocol is of great importance, we offer a secure and practical protocol based on chaotic maps. In this protocol, we take advantage of:  the semi-group property of Chebyshev chaotic map for session key agreement between two participants, the logistic chaotic map for generating non-predictable and pseudo-random sequences, and the Chinese remainder theorem as a one-way theorem. The proposed protocol, in addition to better performance, reduces the setbacks inherent with previous related protocols such as non-anonymity and it does not require the presence of a trusted third party since this element holds the potential to introduce vulnerability. Furthermore, the implementation cost of a trusted third party and complex structure of these protocols are considered the main impediments of utilizing a third party in design and development of key agreement protocols. Thus, the presented protocol has proved to maintain a balance between security and implementation cost.

\section*{References}
\medskip

\bibliographystyle{unsrt}
\bibliography{manuscript3}

\end{document}